\documentclass{jfm}
\usepackage{epsfig}
\usepackage{graphicx}
\usepackage{natbib}
\usepackage{epstopdf}
\usepackage{amsmath}
\usepackage{subfigure}

\ifCUPmtlplainloaded \else
  \checkfont{eurm10}
  \iffontfound
    \IfFileExists{upmath.sty}
      {\typeout{^^JFound AMS Euler Roman fonts on the system,
                   using the 'upmath' package.^^J}%
       \usepackage{upmath}}
      {\typeout{^^JFound AMS Euler Roman fonts on the system, but you
                   dont seem to have the}%
       \typeout{'upmath' package installed. JFM.cls can take advantage
                 of these fonts,^^Jif you use 'upmath' package.^^J}%
      }
  \else
  \fi
\fi

\ifCUPmtlplainloaded \else
  \checkfont{msam10}
  \iffontfound
    \IfFileExists{amssymb.sty}
      {\typeout{^^JFound AMS Symbol fonts on the system, using the
                'amssymb' package.^^J}%
       \usepackage{amssymb}%
         \let\leq=\leqslant
         
      }{}
  \fi
\fi

\ifCUPmtlplainloaded \else
  \IfFileExists{amsbsy.sty}
    {\typeout{^^JFound the 'amsbsy' package on the system, using it.^^J}%
     \usepackage{amsbsy}}
    {\providecommand\boldsymbol[1]{\mbox{\boldmath $##1$}}}
\fi

\newsavebox{\astrutbox}
\sbox{\astrutbox}{\rule[-5pt]{0pt}{20pt}}

\title[Spontaneous generation of inertial waves]{Spontaneous generation of inertial waves from boundary turbulence in a librating sphere}

\author[A. Sauret, D. C\'ebron and  M. Le Bars]%
{Alban Sauret$^1$ \thanks{Email address for correspondence: asauret@princeton.edu}, David C\'ebron$^{1,2}$ and Michael Le Bars$^{1,3}$ }

\affiliation{$^1$ Institut de Recherche sur les Ph\'enom\`enes Hors \'Equilibre, CNRS and Aix-Marseille University, 49 rue F. Joliot-Curie, F-13013 Marseille, France
\\[\affilskip]
$^2$ETH Z\"urich, Institut fur Geophysik, Sonneggstrasse 5, CH8092 Z\"urich, Switzerland
\\[\affilskip]
$^3$Department of Earth and Space Sciences, University of California, Los Angeles, CA 90095-1567, USA}
\pubyear{2013}

\begin{document}

\maketitle

\begin{abstract}
In this work, we report the excitation of inertial waves in a librating sphere even for libration frequencies where these waves are not directly forced. This spontaneous generation comes from the localized turbulence induced by the centrifugal instabilities in the Ekman boundary layer near the equator and does not depend on the libration frequency. We characterize the key features of these inertial waves in analogy with previous studies of the generation of internal waves in stratified flows from localized turbulent patterns. In particular, the temporal spectrum exhibits preferred values of excited frequency. This first-order phenomenon is generic to any rotating flow in the presence of localized turbulence and is fully relevant for planetary applications.
\end{abstract}



\section{Introduction}
Rotating fluids support so-called inertial waves, which are associated with the Coriolis force \cite[][]{kelvin1880}. These waves can be directly excited by various harmonic forcings whose frequencies range between plus and minus twice the spin frequency \cite[][]{greenspanbook}. For instance, \cite{aldridge1969} showed the direct forcing of inertial waves by small periodic oscillations of the spinning rate of a rotating sphere (i.e. "librations") for libration frequencies non-dimensionalized by the mean rotation rate $|\omega_{lib}| \leq 2$ \cite[see also][]{zhang2013}. The nonlinear self-interaction of the excited inertial waves can lead to strong axisymmetric jets, as observed experimentally for tidal forcing \cite[][]{morize2010}. Outside this range of frequencies, inertial waves cannot be excited by direct forcing and were never observed except very recently in an axially librating cylinder in the presence of instabilities near the outer boundary \cite[][]{lopez2011,sauret2012_pof}. The corresponding mechanism is still controversial and is the subject of the present study. We focus here on {{longitudinal}} libration in a spherical geometry, which has recently received renewed interest mainly because of planetary applications \cite[e.g. ][]{rambaux2011}. 

Libration leads to rich dynamics in the contained fluid. At sufficiently large libration amplitudes, centrifugal instabilities are induced near the equator of a sphere where they generate turbulence \cite[][]{noir2009,calkins2010}. This instability is generic to any librating container, as for instance in a cylinder \cite[][]{noir2010,sauret2012_pof}. Libration also induces a mean zonal flow due to nonlinear interactions in the boundary layers even in the absence of inertial waves \cite[][]{busse2010a,sauret2010,noir2012,sauret2012_jfm}. For libration frequency $|\omega_{lib}| \leq 2$, the dynamic of the fluid becomes more complicated as the contribution of the inertial waves to the mean flow is non negligible. The energy fed to inertial waves is then localized near Ekman-layer eruptions \cite[][]{calkins2010,sauret2012_pof,koch2013}. Finally, elliptical instability can also be excited by the libration forcing in ellipsoidal containers \cite[][]{cebronldei}.

In this work, we characterize for the first time the inertial waves excited in a sphere by the boundary flow induced by a longitudinal libration for $\omega_{lib}>2$. In section 2, we present the governing equations and the numerical methods used to tackle this problem. Then section 3 is devoted to the description of the main numerical results. Taking advantage of the well-known similarities between rotating and stratified flows \cite[e.g.][]{veronis1970}, those results are finally explained in section 4 in the form of an analytical model extended from closely related studies of internal waves generation from turbulence \cite[][]{townsend1966,dohan2003,dohan2005,taylor2007}. 

 
\section{Governing equations and numerical methods}

\begin{center}
\begin{figure}
\begin{center}
\includegraphics{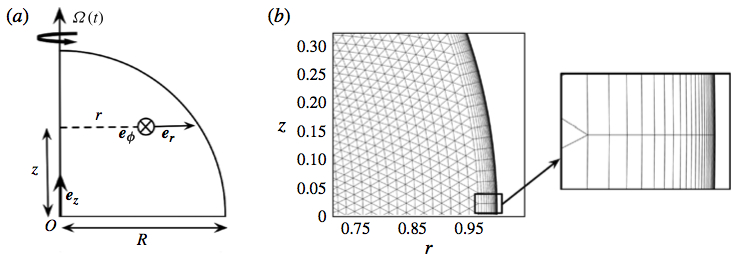}
\end{center}
\caption{(a) Schematic of the upper quarter of the sphere with the cylindrical polar vectors ($\boldsymbol{e_r}$,$\boldsymbol{e_\phi}$,$\boldsymbol{e_z}$). (b) Mesh grid used in the numerical model showing two zones: a bulk zone with triangular elements and a boundary-layer domain with quadrangular elements. {{The inset is a zoom of the mesh grid near the outer boundary}}.}
\label{schematic_sphere}
\end{figure}
\end{center}

Consider the flow in a sphere of radius $R$, filled with an incompressible, homogeneous and Newtonian fluid of kinematic viscosity $\nu$ and density $\rho$, rotating about the $z$-axis at the mean angular velocity $\Omega_0$. In addition to this mean rotation, the sphere oscillates with an angular frequency $\omega_{wall}$ and amplitude $\Delta \Omega$. Using ${\Omega_0}^{-1}$ and $R$ as time and length scales respectively, the instantaneous angular velocity is given by
\begin{equation}
\boldsymbol{\Omega}=[1+\epsilon\,\cos(\omega_{lib}\,t)]\,\boldsymbol{e_z}
\end{equation}
\noindent where $\epsilon=\Delta \Omega/\Omega_0$ and $\omega_{lib}=\omega_{wall}/{\Omega_0}$ are, respectively, the dimensionless libration amplitude and frequency. A schematic of the system is shown in figure \ref{schematic_sphere}(a). The dimensionless equations of motion, written in the rotating frame of reference at the mean angular velocity $\Omega_0$, are
\begin{eqnarray}
\frac{\partial \boldsymbol{u}}{\partial t}+(\boldsymbol{u}\cdot\boldsymbol{\nabla})\,\boldsymbol{u}+2\,\boldsymbol{e_z}\times \boldsymbol{u}&=&-\boldsymbol{\nabla}p+E\,\boldsymbol{\nabla}^2\,\boldsymbol{u}, \\
\boldsymbol{\nabla}\cdot\boldsymbol{u}&=&0,
\end{eqnarray}
\noindent where $\boldsymbol{u}$ and $p$ are, respectively, the velocity field in the rotating frame of reference and the reduced pressure which takes into account the centrifugal force. There are three governing parameters in our problem: the Ekman number, $E=\nu/(\Omega_0\,R^2)$, describing the ratio of the viscous effects and the Coriolis force, $\omega_{lib}$ and $\epsilon$. Throughout this study we use the cylindrical polar unit vectors ($\boldsymbol{e_r}$,$\boldsymbol{e_\phi}$,$\boldsymbol{e_z}$). A no-slip boundary condition is used on the outer boundary where $\boldsymbol{u}=\epsilon\,\cos(\omega_{lib}\,t)\,\boldsymbol{e_z} \times \boldsymbol{r}$.  

An analogous problem of internal wave generation from turbulence in a stratified flow has shown, by comparison with laboratory experiments, that 2D numerical simulations capture the physical mechanism and key features of internal wave generation \cite[][]{dohan2005}. Therefore, to be able to reach sufficiently small Ekman numbers, we base our study on axisymmetric simulations. We use a commercial finite element code, Comsol Multiphysics. This numerical model has already been successfully used to study similar problems of libration-driven flows in spherical geometries \cite[][]{sauret2010,sauret2012_jfm} and cylindrical geometries \cite[][]{sauret2012_pof}. For more details about the numerical procedure and the validation of the numerical code, we refer the reader to \cite{sauret2010}. To ensure a high accuracy, all the simulations used in this work are performed with standard Lagrange elements of $P2-P3$ type (i.e. quadratic for the pressure field and cubic for the velocity field), and the number of degrees of freedom (DoF) used in the simulations is typically of order $350\,000$. A typical mesh grid is shown in figure \ref{schematic_sphere}(b). {{The mesh grid is composed of $17\,794$ elements, of which $6\,500$ are quadrilateral elements devoted to the mesh in the boundary-layer domain and the others are triangular elements for the bulk. More precisely, the boundary-layer domain has a thickness of $0.034$ along the outer boundary and is discretized in the direction normal to the boundary into $25$ quadrilateral elements with an initial thickness of $5.9\,\times \,10^{-5}$ and a stretching factor of $1.2$. This ensures that we have a sufficient number of mesh elements in the viscous layer, typically around $13$ for the parameters considered in this paper. In addition, the adaptive Backward Differentiation Formula (BDF) order is between $1$ and $5$ and the time-step leads to more than $500$ points per period of libration.}}

\section{Numerical results}

\subsection{From stable regime to boundary turbulence}

\begin{center}
\begin{figure}
  \begin{center}
\includegraphics{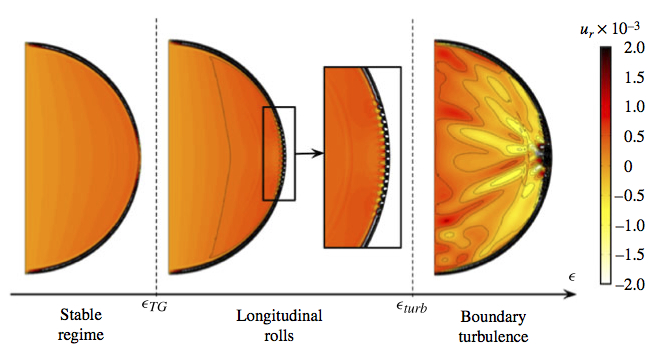}
\end{center}
\caption{Radial velocity $u_r$ in a librating sphere ($\omega_{lib}=3$, $E=5\times 10^{-5}$) for different values of the libration amplitude $\epsilon$ (from left to right, $\epsilon=0.3$, $0.55$, $0.85$); $\epsilon_{TG}$ and $\epsilon_{turb}$ are, respectively, the critical value for the appearance of longitude rolls (see the zoom) and boundary turbulence. Spontaneous generation of inertial waves is observed for $\epsilon > \epsilon_{turb}$.}
\label{diagramme_omega=3}
\end{figure}
\end{center}

In all the following we consider the small Ekman number (${E}\ll1$) and no spin-up regime ($\omega_{lib} \gg \sqrt{E}$), i.e. no spin-up occurs in the bulk at each libration cycle \cite[see e.g.][]{greenspanbook,busse2010a}. This situation is satisfied in planetary fluid layers where the Ekman number is typically smaller than $10^{-12}$ and where the main libration components have frequencies of order $1$ \cite[see e.g.][]{rambaux2011}. As shown in figure \ref{diagramme_omega=3}, three flow regimes are observed in the sphere depending on the libration amplitude $\epsilon$ \cite[see also ][]{noir2009,calkins2010}. For a libration amplitude $\epsilon$ lower than a critical value, called $\epsilon_{TG}$, the flow remains stable and only the viscous layer is visible near the outer boundary; the flow remains laminar everywhere. For $\epsilon > \epsilon_{TG}$ but lower than a critical value, noted $\epsilon_{turb}$, longitudinal rolls, called Taylor-G\"ortler vortices, develop near the equator along the outer boundary. Then, for an amplitude of libration $\epsilon$ larger than the critical amplitude $\epsilon_{turb}$, the longitudinal rolls turn into a turbulent patch localized around the equator. In this last regime, even if the frequency of libration is such that no inertial waves are directly forced (i.e. $\omega_{lib} >2$), inertial waves are excited in the bulk, emitted from the turbulent patch.

\subsection{Characteristics of the generated inertial waves field}

\begin{figure}
\begin{center}
\includegraphics{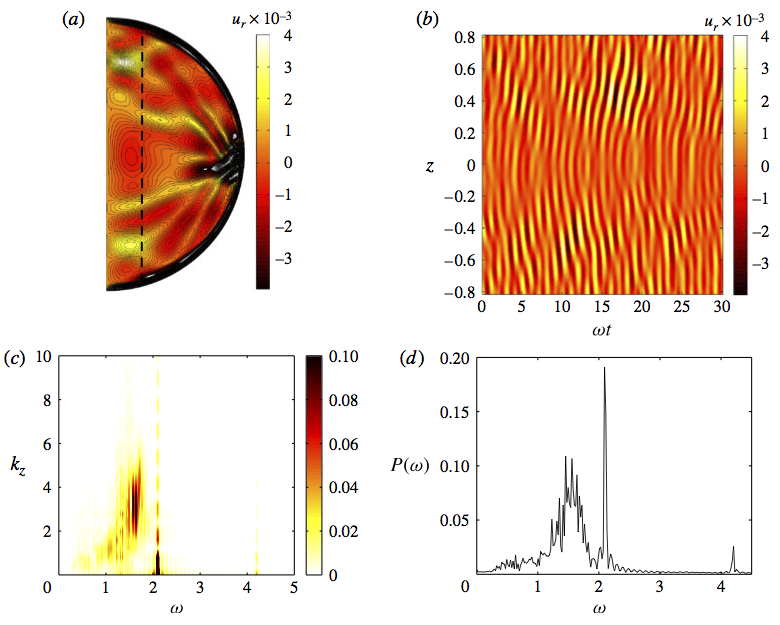}
\end{center}
\caption{Simulations for $\omega_{lib}=2.1$, $E=4\times 10^{-5}$ and $\epsilon=0.8$. (a) Snapshot of the radial velocity field $u_r$ at $t \equiv 5\pi/(4\,\omega_{lib})$. (b) Axial time series of $u_r$ as a function of $z$ taken during $30$ periods at $r=0.2$ and corresponding to the simulation shown in figure \ref{spatio_temporelle}(a) along the dashed line. Apart from the oscillations of the sphere, the propagation of inertial waves is also visible. (c) Corresponding norm of the Fourier transform $P(\omega,k_z)$ showing the axial wavelength and frequency of the generated inertial waves. The colour bar represents $P(\omega,k_z)$. (d) Norm of the Fourier transform of the radial velocity $u_r$ taken at $r=0.2$ and averaged over $z$.}
\label{spatio_temporelle}
\end{figure}

To study the generated inertial waves, we performed series of simulations during $30$ libration periods (once the permanent regime is reached). An example is shown in figure \ref{spatio_temporelle}(a) for $\omega_{lib}=2.1$, $\epsilon=0.8$ and $E=4\times10^{-5}$, together with the space-time diagram of the radial velocity $u_r$ taken at $r=0.2$ in figure \ref{spatio_temporelle}(b). The norm of the two-dimensional Fourier transform of this time series is shown in figure \ref{spatio_temporelle}(c). It exhibits a patch of large amplitude located at $\omega = 2.1$ and small values of $k_z$. This patch is directly related to the libration forcing at the frequency $\omega_{lib} = 2.1$. It is due to the velocity variations associated with the Ekman pumping at each libration cycle. In addition to this main frequency, another patch is observed around $\omega \sim 1.6$ and $k_z \sim 2 - 4$, which is the signature of the inertial waves emitted in the bulk. The norm of the Fourier transform averaged over $z$ for $r=0.2$ is shown in figure \ref{spatio_temporelle}(d). Again, it clearly exhibits a peak located at the libration frequency $\omega=2.1$. In addition, it shows a secondary peak around $\omega=1.4-1.6$ as already observed in the cylinder \cite[see][]{sauret2012_pof}. 

\begin{figure}
  \begin{center}
 \includegraphics{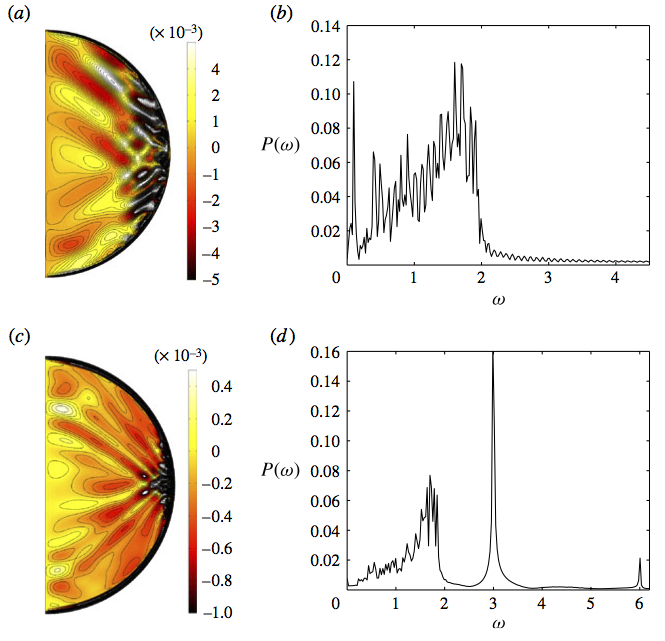}
 \caption{Snapshots of the radial velocity field $u_r$ for $E=4\times 10^{-5}$, $\epsilon=0.8$ at (a) $\omega_{lib}=0.1$ and (c) $\omega_{lib}=3$. Corresponding norms of the Fourier transform of the radial velocity averaged over $z$ for $r=0.2$ at (b) $\omega_{lib}=0.1$ and (d) $\omega_{lib}=3$.} 
    \label{fig:fft_various}
      \end{center}
\end{figure}

We have performed the same analysis for other libration frequencies: $\omega_{lib}=0.1$ (figures \ref{fig:fft_various}a,b) and $\omega_{lib}=3$ (figures \ref{fig:fft_various}c,d). In both cases, the norm of the Fourier transform taken at $r=0.2$ and averaged over $z$ shows a large peak at the libration frequency $\omega= 0.1$ and $\omega=3$, respectively. For $\omega_{lib}=3$, the situation is close to the previous case with a peak around $\omega \sim 1.5$. For $\omega_{lib}=0.1$ a wider range of frequencies is excited in the bulk that may be related to the extended size of the turbulent patch in the Ekman layer. {{Note also that, in this case, the frequency of libration as well as its higher harmonics $2\,\omega_{lib}$, $3\,\omega_{lib}$, ... allow direct forcing of inertial waves. However, the mechanism described above remains fully generic and is superimposed onto these direct forcings, the only difference being that the spectrum is less localized around $\omega \sim 1.4 - 1.6$}}.

\section{Comparison with an idealized analytical model}

\begin{figure}
  \begin{center}
    \includegraphics{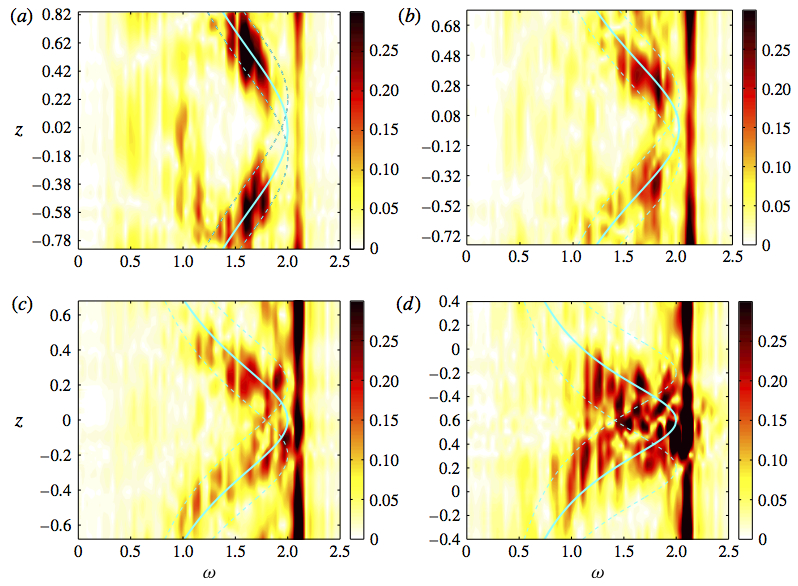}
    \caption{Norm of the Fourier transform $P({\omega})$ as a function of $z$ for slices at different radial positions: (a) $r=0.2$, (b) $r=0.4$, (c) $r=0.6$ and (d) $r=0.8$. The parameters used in the simulations are $\omega_{lib}=2.1$, $\epsilon=0.7$ and $E=4\times10^{-5}$. The continuous cyan lines represent the geometrical path given by relation (\ref{IW_pos_theo}), and the dashed cyan lines take into account the axial extension of the turbulent patch following (\ref{relation_ff}). } 
    \label{fig:fft_rrrr}
      \end{center}
\end{figure}

Following the seminal study of \cite{townsend1966} and the more recent work of \cite{taylor2007} for internal waves, the inertial waves propagation in the bulk of the sphere can be understood by a linear model. For the sake of simplicity, we consider here a two-dimensional model and neglect the curvature of the boundary of the sphere. In his study of internal waves generated at a horizontal interface, \cite{townsend1966} considered a gaussian perturbation in both space and time along the horizontal plane $z=0$ of the vertical displacement, which is directly related to the vertical velocity. Following the analogy between stratified and rotating flows, it is natural to consider here a perturbation of the radial velocity along the plane $r=1$ and localized around the equator $(r=1, z=0)$.  To perform a tractable analysis, we assume this perturbation to be separable in space and time, i.e.
\begin{equation}\label{TT1}
u_r^{p}(z,t)=f(z)\,g(t),
\end{equation}
where the spatial function $f(z)$ is maximum at $z=0$. In the following, owing to the symmetry of the system with respect to the equator of the sphere, we only consider the half-domain $z>0$. The Fourier transform of the expression (\ref{TT1}) leads to
\begin{eqnarray}
\hat{u}_r^{p}(k_z,{\omega})& =&\int \int u_r^{p}(z,t)\,\exp\left[-\text{i}\,k_z\,z+\text{i}\,{\omega}\,t\right] \text{d}t \,\text{d}z\, = \hat{f}(k_z)\,\hat{g}({\omega}) \label{eq:cotcot}
\end{eqnarray}
where $\mathbf{k}=(k_r,k_z)$ is the wavevector and ${\omega}$ is the frequency. The propagation of this perturbation along the r-direction can be written as :
\begin{equation}
u_r(r,z,t)=\frac{1}{4\,\pi^2}\,\int \int \hat{f}(k_z)\,\hat{g}({\omega})\,\text{exp}\left[\text{i}\,(k_r\,(1-r)+k_z\,z-{\omega}\,t)\right] \text{d}{\omega}\, \text{d}k_z
 \label{eq:cotcotback1}
\end{equation}
The dispersion relation for inertial waves links the axial and radial components of the wavevector:
\begin{equation}
{k_r}=\pm {k_z}\,\sqrt{\frac{4}{{\omega}^2}-1} \label{gen_omega}
\end{equation}
The radial velocity field then can be written as
\begin{equation}
u_r(r,z,t)=\frac{1}{4\,\pi^2}\,\int_{-2}^2\,\hat{g}({\omega})\,\text{e}^{-\text{i}{\omega} t}\,\text{d}{\omega}\,\int \hat{f}(k_z)\,\text{exp}\left[\text{i}\,k_z\left(z\pm (1-r)\,\sqrt{\frac{4}{{\omega}^2}-1}\right)\right] \text{d}k_z,
 \label{eq:cotcotback}
\end{equation}
neglecting the contribution for $|{\omega}|>2$, which corresponds to evanescent waves \cite[][]{greenspanbook}. The relation (\ref{eq:cotcotback}) can be rewritten as
\begin{equation}
u_r(r,z,t)=\frac{1}{2\,\pi}\,\int_{-2}^2\,\hat{g}({\omega})\,f\!\left(z\pm (1-r)\,\sqrt{\frac{4}{{\omega}^2}-1}\right)\,\text{e}^{-\text{i}{\omega} t}\,\text{d}{\omega}
 \label{eq:cotcotback2}
\end{equation}
The temporal Fourier transform of the radial velocity at a position ($r$, $z$) is thus
\begin{equation}
\hat{u}_r(r,z,{\omega})=\hat{g}({\omega})\,f\!\left(z\pm (1-r)\,\sqrt{\frac{4}{{\omega}^2}-1}\right)
 \label{eq:cotcotback3}
\end{equation}
for $|{\omega}| \leq 2$, with a negligible contribution from $|{\omega}|>2$. This expression shows that the signal in the sphere, at first order and neglecting the viscosity, is the product of the temporal signal of the initial excitation $\hat{g}({\omega})$ and the spatial function $f$. Remembering that this spatial function is localized around $0$, the energy in the $z>0$ space is mainly focused around trajectories
\begin{equation} \label{IW_pos_theo}
z-(1-r)\,\sqrt{\frac{4}{{\omega}^2}-1}=0, \mbox{\quad i.e \quad} {\omega}=\frac{2\,(1-r)}{\sqrt{(1-r)^2+z^2}}.
\end{equation}
This is readily interpreted geometrically. Starting from a point source of fluctuations, waves of all frequencies are emitted with a given amplitude, which depends on the initial excitation in the turbulent patch. Waves with $|{\omega}| > 2$ are evanescent, while waves with $|{\omega}| \leq 2$ propagate following the dispersion relation ${\omega}=2\,\cos\theta$, where $\theta$ is the angle of propagation \cite[see e.g.][]{greenspanbook}. This means that a given point $(r,z)$ can only be reached by a wave having the specific frequency given by (\ref{IW_pos_theo}). 
This geometrical path is plotted in figure \ref{fig:fft_rrrr}(a)-(d) for different radial positions and shows good agreement with the numerical results.

\begin{figure}
  \begin{center}
   \includegraphics{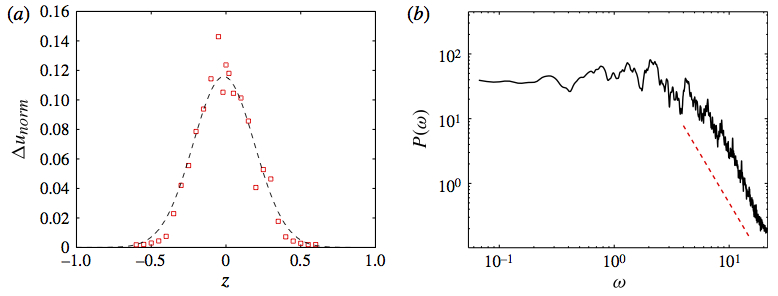}
      \caption{(a) Maximum fluctuation of the velocity normal to the outer boundary as a function of $z$. Parameters are: $\epsilon=0.8$, $E=4\times10^{-5}$ and $\omega_{lib}=2.1$. Red squares are the numerical results and the black dashed line is a gaussian fit centered at the equator and of standard deviation $\sigma = 0.2$. (b) Norm of the Fourier transform $P({\omega})$ of the fluctuation of the radial velocity $u_r$ close to the outer boundary ($r=0.98,\,z=0$). Parameters are the same as panel (a). The dashed red line shows the slope ${\omega}^{-3}$. } 
    \label{fig:exict_initial}
      \end{center}
\end{figure}

To go further, we need the two functions $f$ and $g$, which can be obtained from simulations. The axial extension of the source, $f(z)$, is well modelled by a gaussian fit, as illustrated in figure \ref{fig:exict_initial}(a), where we have plotted the maximum amplitude of the fluctuation of the velocity normal to the outer boundary, i.e. the spherical radial velocity.  {{Note that considering the time-averaged velocity normal to the outer boundary leads to a similar lateral extension of the turbulent patch.  }}
Taking into account this extension of the source allows one to estimate the extension of the propagating signal around the geometrical path (\ref{IW_pos_theo}), taking for instance
\begin{equation} \label{relation_ff}
z-(1-r)\,\sqrt{\frac{4}{{\omega}^2}-1}= \pm \sigma,
\end{equation}
where $\sigma$ is the standard deviation of the gaussian fit. Results are shown in figure \ref{fig:fft_rrrr}(a)-(d) and show good agreement with the numerical results.

\begin{figure}
  \begin{center}
    \includegraphics{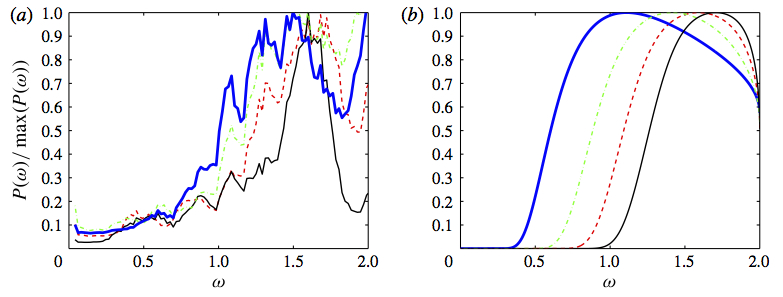}
          \caption{Rescaled norm of the Fourier transform $P({\omega})/\text{max}\left(P({\omega})\right)$ for ${\omega} \in[0,\,2]$ of the fluctuation of $u_r$ at different radial locations averaged over $z$. (a) Numerical results {{for $\omega_{lib}=2.1$}} and (b) analytical model plotted with the relation (\ref{eq:cotcotback398}). We have reported the norm of the Fourier transform at $r=0.8$ (blue thick line), $r=0.6$ (green dash-dotted line), $r=0.4$ (red dashed line) and $r=0.2$ (black continuous line).} 
    \label{cotcotback3982}
      \end{center}
\end{figure}

A typical temporal Fourier transform of the excitation is shown in figure \ref{fig:exict_initial}(b). At large frequencies, the turbulent patch is well fitted by a slope ${\omega}^{-3}$ characteristic of a two-dimensional turbulence \cite[see e.g.][]{smith,bofetta} For ${\omega} \in [0,2]$ corresponding to the propagative waves, the norm of the Fourier transform is nearly constant; hence according to (\ref{eq:cotcotback3}), no frequency is a priori preferred in the bulk. Yet, as seen before, the numerical estimation of the norm of the Fourier transform averaged over $z$ exhibits a maximum, whose location depends on the radial position (see fig. \ref{cotcotback3982}). This focalization was also observed in a librating cylinder \cite[][]{sauret2012_pof} for a turbulent patch localized all along the outer boundary and for the generation of internal waves from a turbulent layer \cite[][]{dohan2005,taylor2007}. In this latter case, the observed frequency selection was attributed to the influence of viscosity. Adapting their study to inertial waves, the viscous attenuation of a given wave with axial wavenumber $k_z$ and frequency ${\omega}$ propagating in the $r$-direction, is given by\begin{equation} \label{ohayoo}
\text{exp}\left[ -\frac{16\,E\,|{k_z}|^3}{{\omega}^4\,\sqrt{4-{\omega}^2}}\,(1-r)\right],
\end{equation}
which, according to the typical values considered here (see e.g. figure \ref{spatio_temporelle}(c)), induces a change in the wave amplitude of less than $0.01\%$. Viscous attenuation can clearly not account for the frequency selection shown here, and this will be even more true in planetary applications where $E < 10^{-12}$. In the present case, the finite size and the shape of the spherical container are actually responsible for the selection of preferred frequencies. Indeed, using the numerically determined values of the source functions $f$ and $g$, we can average over $z$ the signal given by the relation (\ref{eq:cotcotback3}) at a constant $r$ between the boundaries of the sphere $z=\pm \sqrt{1-r^2}$, which gives
\begin{eqnarray}
<\hat{u}_r(r,{\omega})>_z &  \propto & \left(\text{erf}\left[\frac{(1-r)\, |k_r/k_z|}{\sigma\,\sqrt{2}}\right]-\text{erf}\left[\frac{(1-r)\,|k_r/k_z|-\sqrt{1-r^2}}{\sigma\,\sqrt{2}}\right]\right)\,\hat{g}({\omega}), \nonumber 
\label{eq:cotcotback398} \\
\end{eqnarray}
where $\text{erf}$ is the Gauss error function and $|k_r/k_z|=\sqrt{{4}/{{\omega}^2}-1}$ (see eq. \ref{gen_omega}). Figure \ref{cotcotback3982} shows the comparison between numerical and analytical results at different radial positions. It exhibits reasonable agreement, taking into account the strong assumptions used in the analytical model. {{Starting from an excitation localized at the equator and with a flat spectrum between $[0,\,2]$, energy is emitted in all directions. But because of geometrical constraints, energy propagation along or close to the direction of the axis of rotation, corresponding to low-frequency waves, is more prevented than propagation along the equator, corresponding to frequencies close to 2. This geometrical frequency selection has a growing importance while going further away from the source, i.e. while $r$ decreases. Hence, while $r$ decreases, selected frequencies are more and more localized close to 2.}}
{{Note also that the mismatch between the numerical results and the analytical model close to $\omega \sim 2.0$ can be explained by our assumptions. The analytical model is based on a flat excited spectrum in the range $\omega\in[0,\,2]$ only. In the numerical simulation, a peak around the frequency $\omega=\omega_{lib}=2.1$ is also present,  associated with the spin-up and spin-down close to the outer boundary at each libration cycle (see also Fig. 3).}}

\section{Conclusion}
In conclusion, this work presents the first evidence of spontaneous generation of inertial waves from a localized patch of turbulence in a rotating container and of the related mechanism of frequency selection by geometrical constraints. These processes are illustrated here by a simple two-dimensional analytical model and by simple axisymmetric simulations of a librating sphere. This configuration is especially interesting for planetary applications: celestial bodies such as Io and Europa, are indeed thought to present turbulence around their equator driven by libration \cite[][]{noir2009}. The present mechanism may then participate in the explanation of their preferred eigenmodes of vibration. More generally, we expect the processes presented here to be fully generic to any source of turbulence and to any type of container.  A difference must be made between localized  and extended patches, as illustrated by the libration of a sphere and of a cylinder for inertial waves. Following the study of \cite{townsend1966} for internal waves \cite[see also ][]{taylor2007}, the resulting signal for an extended source is equal to the superposition of the effects of random localized sources in space and time. In all cases, a focusing of energy at a given frequency is thus to be expected.

\section*{Acknowledgments}

D.C. acknowledges support from the ETH Z\"urich Postdoctoral Fellowship Progam and from the Marie Curie Actions for People COFUND Program. M.L.B. acknowledges support from the Marie Curie Actions of the European Commission (FP7-PEOPLE-2011-IOF).

\bibliographystyle{jfm}
\bibliography{biblio}

\end{document}